\documentclass[twocolumn,showpacs,preprintnumbers,amsmath,amssymb]{revtex4}
\usepackage{graphicx}
\usepackage{dcolumn}
\usepackage{bm}

\newcommand{\ha}{\hat{a}}

\newcommand{\hAD}{\hat{\cal A}}

\newcommand{\hap}{\hat{a}^+}

\newcommand{\hApD}{\hat{\cal A}^+}

\newcommand{\hH}{\hat{H}}
\newcommand{\hHD}{\hat{\cal H}}

\begin{document}
\title{Trembling motion of relativistic electrons in a magnetic field}
\date{\today}
\author{Tomasz M. Rusin*}
\author{Wlodek Zawadzki\dag}
\affiliation{*PTK Centertel Sp. z o.o., ul. Skierniewicka 10A, 01-230 Warsaw, Poland; email:tmr@vp.pl\\
            \dag Institute of Physics, Polish Academy of Sciences, Al. Lotnik\'ow 32/46, 02-688 Warsaw, Poland}

\begin{abstract}
Zitterbewegung (ZB, the trembling motion) of free relativistic electrons in a vacuum in the presence of
an external magnetic field is calculated. It is shown that the motion of an electron wave packet has
intraband frequency components, corresponding to the classical cyclotron motion, and several interband frequency
components corresponding to the Zitterbewegung. For a two-dimensional situation, the presence of a magnetic field
makes the ZB motion stationary, i.e. not decaying in time. We show how to simulate the ZB in a magnetic
field using trapped ions and laser excitations in the spirit of recently observed proof-of-principle
ZB simulation by Gerritsma {\it et al.} Nature {\bf 463}, 68 (2010). It is demonstrated that,
for the parameters of the Dirac equation simulated by the above experiment, the effect of a magnetic field
on the Zitterbewegung is considerable.
\end{abstract}

\maketitle

The phenomenon of Zitterbewegung was theoretically devised by Schrodinger~\cite{Schroedinger1930}
for free relativistic
electrons in a vacuum. Schrodinger showed that, due to a noncommutativity of the quantum velocity operators
with the Dirac Hamiltonian, relativistic electrons experience the trembling motion even in
absence of external potentials. Here one deals with a purely quantum effect since it goes beyond Newton's
first law of classical motion. It was later recognized that the phenomenon of ZB is due to an interference of
electron states with positive and negative electron energies~\cite{BjorkenBook,ThallerBook}.
The predicted frequency of ZB oscillations is
very high, corresponding to $\hbar\omega_Z \simeq 2mc^2$, and its amplitude is very small being around the Compton
wavelength $\hbar/mc = 3.86\times$ 10$^{-3}$~\AA.
Thus, it is impossible to observe this effect using  present experimental arrangements.
In fact, even the principal observability of ZB was often questioned in the literature,
see e.g.~\cite{Huang1952,Krekora2004}.
However, in a very recent paper Gerritsma {\it et al.}~\cite{Gerritsma2010} simulated the 1+1 Dirac equation
(DE) and the resulting Zitterbewegung with the use of trapped Ca ions excited by
appropriate laser beams. The remarkable advantage of this method is that one can simulate the
basic parameters of DE, i.e. $mc^2$ and $c$, and give them desired values. This results in a much lower
ZB frequency and a much larger ZB amplitude. These were in fact observed.

The purpose of our work is twofold. First, we calculate the Zitterbewegung of relativistic
electrons in a vacuum in the presence of an external magnetic field.
The presence of a constant magnetic field
does not cause electron transitions between negative and positive electron energies. On the other hand,
it quantizes the energy spectrum into Landau levels which brings qualitatively new features into the ZB.
The problem of ZB in a magnetic field was treated
before~\cite{Barut1985}, but the results were limited to the operator level and suffered
from various deficiencies which we briefly mention below.
A similar problem was treated in Ref.~\cite{Villavicencio2000} on the operator level
in weak magnetic field limit.
Our second purpose is to show how to simulate the Dirac equation
for electrons in a magnetic field with the use of trapped ions. We demonstrate that, using  the
parameters of DE simulated in Ref.~\cite{Gerritsma2010}, the effects of the simulated magnetic field  on ZB
are sufficiently strong to be observable.

The Hamiltonian for a relativistic electron in a magnetic field is
\begin{equation} \label{GenHD}
{\cal \hat{H}} = c\alpha_x\hat{\pi}_x + c\alpha_y\hat{\pi}_y + c\alpha_z\hat{\pi}_z + \beta mc^2,
\end{equation}
where $\hat{\bm \pi} = \hat{\bm p}- q\bm A$ is the generalized momentum, $q$
is the electron charge, $\alpha_i$ and $\beta$ are the Dirac matrices in the standard notation.
Taking a magnetic field $\bm B \| \bm z $ we choose the vector potential
$\bm A = (-By,0,0)$ and look for eigenfunctions in the form
\begin{equation} \label{GenPsi}
\Psi(\bm r) = e^{ik_xx+ik_zz}\Phi(y).
\end{equation}
Introducing the magnetic radius $L = \sqrt{\hbar/eB}$ and
$\xi=y/L-k_xL$, we have $y=\xi L  +  k_xL^2$, $eBy/\hbar=y/L^2$,
and $\partial/ \partial y = (1/L)\partial/ \partial \xi$.
Defining the standard raising and lowering operators for the harmonic oscillator
$\ha_y =(\xi+ \partial/\partial \xi )/\sqrt{2}$ and
$\hap_y= (\xi -\partial/\partial \xi)/\sqrt{2}$
one has $[\ha_y,\hap_y]=1$ and $\xi = (\ha_y + \hap_y)/\sqrt{2}$. Now the Hamiltonian reads
\begin{equation} \label{GenDHH}
 {\cal \hat{H}} = \left(\begin{array}{cc}  mc^2\hat{\bm 1}   & \hH   \\
    \hH  & -mc^2 \hat{\bm 1} \\  \end{array} \right),
\end{equation}
where $\hat{\bm 1}$ is the 2$\times$2 identity matrix, and
\begin{equation} \label{GenHaap}
 \hH = -\hbar\omega\left(\begin{array}{cc}  cp_z & \ha_y \\ \hap_y & -cp_z \\  \end{array}\right),
\end{equation}
with $\omega=\sqrt{2}c/L$. An eigenstate $|{\rm n}\rangle$ of $\hH$ is
characterized by five quantum numbers: $n,k_x,k_z,\epsilon,s$,
where $n$ is the Landau level number, $k_x$ and $k_z$ are the wave vector components,
$\epsilon=\pm 1$ labels the positive and negative energy branches,
and $s=\pm 1$ is the spin index.
In the representation of  Johnson and Lippmann~\cite{Johnson1949} the state $|{\rm n}\rangle$ is
\begin{equation} \label{BJohnson}
|{\rm n}\rangle =  \frac{e^{ik_xx+ik_zz}}{N_{n\epsilon k_z}}\left(\begin{array}{rl}
   s_u(\epsilon E_{nk_z}+mc^2)&|n-1\rangle \\
   s_l(\epsilon E_{nk_z}+mc^2)&|n\rangle \\
  (s_u\hbar k_zc -s_l\hbar \omega_n)&|n-1\rangle   \\
  -(s_u\hbar \omega_n + s_l \hbar k_zc)&|n\rangle \end{array}\right),
\end{equation}
where $s_u=(s+1)/2$ and $s_l=(s-1)/2$ are the projection operators on the states $s=\pm 1$, respectively.
The frequency is $\omega_n=\omega \sqrt{n}$, the energy is
\begin{equation} \label{BEnkz}
 E_{nk_z}=\sqrt{(mc^2)^2 + (\hbar\omega_n)^2 + (\hbar k_zc)^2 },
\end{equation}
and the norm is $N_{n\epsilon k_z}=(2E_{nk_z}^2+2\epsilon mc^2E_{nk_z})^{1/2}$.
In this representation the energy $E_{nk_z}$ does not depend explicitly on $s$.
The harmonic oscillator states $\langle \bm r |n\rangle$ have the standard form.

In order to calculate the average electron position we introduce
four-component operators $\hAD= \textrm{diag}(\ha_y,\ha_y,\ha_y,\ha_y)$
and $\hApD= \textrm{diag}(\hap_y,\hap_y,\hap_y,\hap_y)$.
Now we define $\hat{\cal Y}= L(\hAD + \hApD)/\sqrt{2}$ and $\hat{\cal X}= L(\hAD - \hApD)/i\sqrt{2}$
in analogy to the position operators $\hat{y}$ and $\hat{x}$.
We use averaging of $\hAD$ and $\hApD$ operators in the Heisenberg picture
$\hAD(t)=e^{i\hH t/\hbar}\hAD(0)e^{-i\hH t/\hbar}$ over a wave packet $f(\bm r)$. Inserting
two unity operators $\hat{\bm  1} = \sum_{\rm n}|{\rm n}\rangle\langle {\rm n}|$ one obtains
\begin{eqnarray} \label{ADt0}
\langle \hAD(t)\rangle =
 \sum_{\rm nn'}   \langle f| {\rm n}\rangle  e^{i\epsilon E_{\rm n}t/\hbar }
   \langle {\rm n}|\hAD|  {\rm n'} \rangle
  e^{-i\epsilon' E_{\rm n'}t/\hbar}\langle {\rm n'}|f\rangle,
\end{eqnarray}
and similarly for $\langle \hApD(t)\rangle$.
The wave packet is assumed to be in the form $f(\bm r)^i = f^i_z(z)f^i_{xy}(x,y)$, where $i=1,2,3,4$
labels the packet components. Below we limit our calculations to a packet with the second nonzero
component. The summation over ${\rm n},{\rm n'}$ denotes summations over
$n,n',\epsilon,\epsilon',s,s'$ and integrations over $dk_x,dk^{'}_{x},dk_z,dk^{'}_z$. We obtain explicitly
\begin{eqnarray} \label{ADt1}
\langle \hAD(t)\rangle =
\sum_{n,n'}\int_{-\infty}^{\infty} dk_z dk^{'}_z  g^i_z(k_z) g^i_z(k^{'}_z) \times \nonumber \\
\sum_{\epsilon,\epsilon'}
e^{i(\epsilon E_{nk_z}-\epsilon' E_{n'k^{'}_z})t/\hbar}  \chi_{n\epsilon k_z}\chi_{n^{'}\epsilon^{'} k^{'}_z}
  \times \nonumber \\
\sum_{s,s'} \int_{-\infty}^{\infty} dk_x dk^{'}_x
 s_l s_l' F^{2*}_{n}(k_x) F^2_{n'}(k^{'}_x)  \langle {\rm n}|\hAD|  {\rm n'} \rangle,
\end{eqnarray}
where $\chi_{n\epsilon k_z} =(\epsilon E_{nk_z}+mc^2)/N_{n\epsilon k_z}$, and
\begin{equation} \label{ADFj}
  F^j_n(k_x) = \frac{1}{\sqrt{2L}C_n} \int_{-\infty}^{\infty} g^j_{xy}(k_x,y)e^{-\frac{1}{2}\xi^2}{\rm H}_{n}(\xi)dy,
\end{equation}
in which
\begin{equation} \label{ADgjxy}
 g^j_{xy}(k_x,y) = \frac{1}{\sqrt{2\pi}} \int_{-\infty}^{\infty}  f_{xy}^j(x,y)e^{ik_xx} dx,
\end{equation}
and
\begin{equation} \label{ADgjz}
 g_z^j(k_z) = \frac{1}{\sqrt{2\pi}} \int_{-\infty}^{\infty}  f_{z}^j(z)e^{ik_zz} dz.
\end{equation}
The selection rules for $\langle {\rm n}|\hAD|  {\rm n'} \rangle$ are
$n'=n+1$, $k_x'=k_x$, $k_z'=k_z$, while for $\langle {\rm n}|\hApD| {\rm n'} \rangle$
they are $n'=n-1$, $k_x'=k_x$, $k_z'=k_z$.
After some manipulation we finally obtain
\begin{eqnarray} \label{ADsum}
\langle\hAD(t)\rangle\!\!  &=& \frac{1}{2}\sum_n \sqrt{n+1} U_{n,n+1}^{2,2} \left(I^+_c +  I^-_c + iI^+_s + iI^-_s \right),\ \ \ \ \ \  \\
\langle\hApD(t)\rangle\!\! &=& \frac{1}{2}\sum_n \sqrt{n+1} U_{n+1,n}^{2,2} \left(I^+_c +  I^-_c - iI^+_s - iI^-_s \right),\ \ \ \ \ \
\end{eqnarray}
where
\begin{eqnarray} \label{ADIcs}
 I^{\pm}_c &=& \int_{-\infty}^{\infty}\left(1 \pm \frac{E_{nk_z}}{E_{n+1,k_z}} \right)|g^j_z(k_z)|^2 \times \nonumber \\
           && \ \ \ \ \ \ \cos\left[(E_{n+1,k_z} \mp E_{nk_z})t/\hbar \right]dk_z,  \\
 I^{\pm}_s &=&  mc^2\int_{-\infty}^{\infty} \left(\frac{1}{E_{nk_z}} \pm \frac{1}{E_{n+1,k_z}} \right)
           |g^j_z(k_z)|^2 \times \nonumber \\
           && \ \ \ \ \ \ \sin\left[(E_{n+1,k_z} \mp E_{nk_z})t/\hbar \right]dk_z,
\end{eqnarray}
and
\begin{equation} \label{AUmn}
 U^{i,j}_{m,n} = \int_{-\infty}^{\infty} F_{m}^{i*}(k_x) F_{n}^{j}(k_x) dk_x,
 \end{equation}
for $i,j =1,2,3,4$. We perform specific calculations taking $f(\bm r)$ in the form of an
ellipsoidal Gaussian packet $(0,f(\bm r),0,0)$
characterized by three widths $d_x$, $d_y$, $d_z$ and having a non-zero momentum
$\hbar \bm k_0=\hbar(k_{0x},0,0)$.
The quantities $g_{xy}(k_x,y)$, $F^{i}_n(k_x)$, and $U_{m,n}^{i,j}$
can be obtained analytically, see Ref.~\cite{Rusin20008a}.

\begin{figure}
\includegraphics[width=8.5cm,height=5.5cm]{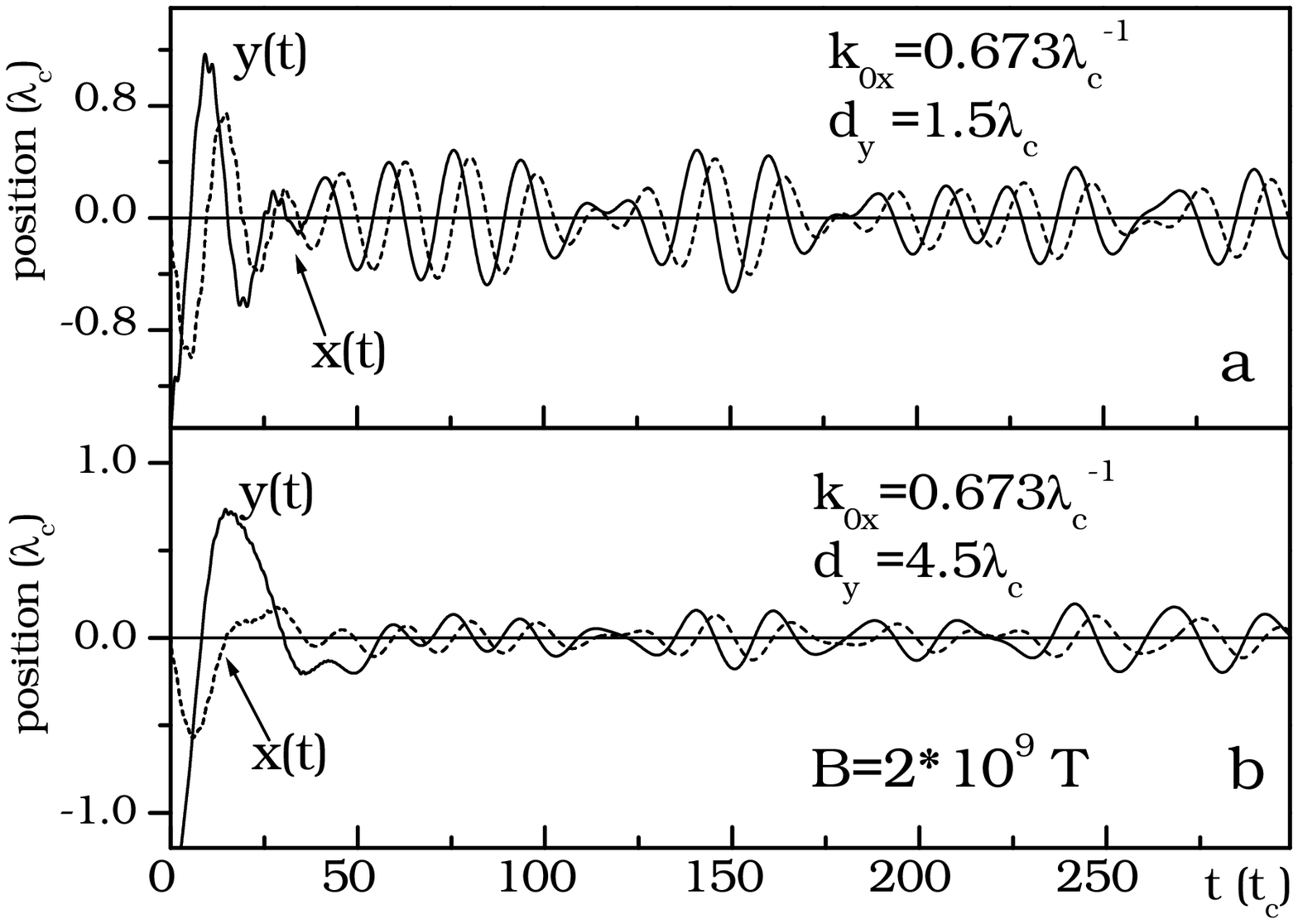}
\caption{Calculated motion of the electron wave packet in vacuum in the  presence of a gigantic
         magnetic field $B=2\times$10$^9$~T according to 3+1 Dirac equation for two packet widths.
         Both interband and intraband frequencies are present in the spectrum. Time is given in
         $t_c=\hbar/mc^2$, positions are given in Compton wavelengths
         $\lambda_c=\hbar/mc$. } \label{FigC}
\end{figure}

In the non-relativistic limit: $\hbar\omega, cp_z \ll mc^2$,  equations~(\ref{ADsum})$~-~$(\ref{AUmn})
reduce to the cyclotron motion with the frequency
$\omega_c=(E_{n+1,k_z}-E_{n,k_z})/\hbar\simeq eB/m$ and the radius  $r=k_{0x}L^2=mv/eB$.
In this limit the ZB part of the motion has the amplitude several orders of magnitude
smaller than $\lambda_c$. In the opposite
relativistic limit, in which $\hbar\omega, cp_z \simeq mc^2$,
both the cyclotron and ZB components have comparable amplitudes.
In Fig.~\ref{FigC} we plot the average position in $x$ and $y$ directions for an electron wave
packet in a gigantic magnetic field $B=2\times$10$^9$~T for two packet parameters.
It is seen that the ZB oscillations consist of several frequencies. This is the main effect of the
magnetic field, which quantizes both positive and negative electron energies into Landau levels.
In larger time scale the oscillations in 3+1 space go through decays and revivals, but finally disappear.

The involved DE must have at least 2+1 dimensions since a magnetic field $\bm B \parallel \bm z$ couples the
electron motion in $\bm x$ and $\bm y$ directions.
The above equations, derived for the 3+1 DE,  can be reduced to the 2+1 case by
setting $|g_j(k_z)|^2=\delta(k_z)$.
The main difference between 3+1 and 2+1 cases is that for the 3+1 case the ZB of the
wave packet has a transient character (it decays in time), whereas in the 2+1 case it has a
permanent character going through decays and revivals. This difference, due to the integration
over $k_z$ in the 3+1 case, is in agreement with general predictions
of Lock~\cite{Lock1979}. However, a slow decay of ZB in time will occur also in the 2+1 case
since the trembling electron will emit radiation. This is possible because a Gaussian wave packet
is not an eigenstate of the Dirac Hamiltonian. Also a broadening of Landau levels due to external
perturbations results in a transient character of ZB, c.f.~\cite{Rusin2009}.

The main experimental problem in investigating the above ZB phenomenon in an external magnetic field
is the fact that for free relativistic electrons the basic ZB (interband) frequency corresponds to
the energy $\hbar\omega_Z \simeq 1$~MeV, whereas the cyclotron frequency for a magnetic
field of 100~T is $\hbar\omega_c \simeq 0.1$~eV, so that the magnetic effects in ZB are
very small. However, as mentioned above, it is possible by now to simulate
the Dirac equation changing thereby its basic parameters. This gives a possibility
to modify drastically the ratio $\hbar\omega_c/2mc^2$ making its value much more advantageous.
To simulate the DE for an electron in a magnetic field, we transform it first to the form
$\hHD^{'}= c\sum_{i}\alpha_i\hat{p}_i+\delta mc^2$, using the unitary operator
$\hat{P}=\delta(\delta+\beta)/\sqrt{2}$, where
$\delta=\alpha_x\alpha_y\alpha_z\beta$~\cite{Moss1976}. After the transformation the Hamiltonian is
$\hHD^{'}=\left(\begin{array}{cc} 0 &\hH^{'} \\ \hH^{'\dagger} &0 \end{array} \right)$,
where
\begin{equation} \label{IonHD}
\hH^{'} = \left(\begin{array}{cc} c\hat{p_z} - imc^2 &c\hat{p_x}-\hbar\omega \ha_y \\
     c\hat{p_x}-\hbar\omega \hap_y &-c\hat{p_z} - imc^2 \end{array} \right).
\end{equation}
We then follow the approach of Lamata {\it et al.}~\cite{Lamata2007}
and consider a four-level system of Ca or Mg ions. Simulations of $cp_x$ and $cp_z$
terms in the above Hamiltonian  are realized the same way as for a free Dirac particle.
Thus one uses pairs of the Jaynes-Cumminngs (JC) interaction:
              $\hH^{\phi_r}_{JC}= \hbar\eta\tilde{\Omega}(\sigma^+\ha  e^{i\phi_r}+\sigma^-\hap e^{-i\phi_r} )$,
and the anti-Jaynes-Cumminngs (AJC) interaction:
              $\hH^{\phi_b}_{AJC}= \hbar\eta\tilde{\Omega}(\sigma^+\hap e^{i\phi_b}+\sigma^-\ha  e^{-i\phi_b})$,
while a simulation of $mc^2$ is done by the carrier interaction
$\hH^{c}= \hbar\Omega(\sigma^+e^{i\phi_c}+\sigma^-e^{-i\phi_c})$. Here $\Omega$ and $\tilde{\Omega}$ are
the coupling strengths and $\eta$ is the Lamb-Dicke parameter.
On the other hand, a simulation of $\ha_y$ and $\hap_y$ (which include the magnetic field)
can be  done by single JC or AJC interactions. Within the notation of
Refs.~\cite{Lamata2007,Johanning2009,Leibfried2003}
one may simulate $\hHD^{'}$ by the following set of excitations
\begin{eqnarray}  \label{IonSim}
\hHD^{'}_{ion} &=& \hH^{p_x}_{\sigma_x (ad)}+ \hH^{p_x}_{\sigma_x (bc)}+
               \hH^{\phi_r=\pi}_{JC (ad)}  + \hH^{\phi_b=\pi}_{AJC (bc)}  + \nonumber \\
&&             \hH^{p_z}_{\sigma_x (ac)}- \hH^{p_z}_{\sigma_x (bd)}+
               \hH^{c}_{\sigma_y (ac)}+ \hH^{c}_{\sigma_y (bd)},
\end{eqnarray}
where $\hH^{p_q}_{\sigma_j}= 2\eta_q\tilde{\Omega}\sigma_j \Delta_q p_q$, $p_q = i\hbar(\hap_q-\ha_q)/\Delta_q$,
$j,q=x,z$, and the Pauli matrices $\sigma_j$ are obtained from combinations
of $\hH^{\phi_r}_{JC}$ and $\hH^{\phi_b}_{AJC}$ with appropriate
phases $\phi_r$ and $\phi_b$. Here the spread of the ground ion wave function is $\Delta_q=\sqrt{\hbar/2M\nu_q}$
and the Lamb-Dicke parameter is $\eta_q=k\sqrt{\hbar/2M\nu_q}$, where
$M$ is ion's mass, $\nu_q$ is trap's frequency in the $\bm q$ direction and
$\bm k$ is the  wave vector of the driving field in a trap. The subscripts in parenthesis of Eq.~(\ref{IonSim})
denote states involved in a given  transition. The JC interaction gives $\ha_y$ in $\hH^{'}_{12}$
and $\hap_y$ in $\hH^{'\dagger}_{21}$
elements of the Hamiltonian $\hH^{'}$ in Eq.~(\ref{IonHD}), while AJC gives $\ha_y$ in
$\hHD^{'}_{21}$ and $\hap_y$ in  $\hH^{'\dagger}_{12}$ elements.
The simulation of 3+1 DE by Eq.~(\ref{IonSim}) can be realized with 12 pairs of laser excitations.
If one omits the $p_z$ interaction, which corresponds to the 2+1 DE, one needs 8 pairs of laser excitations.
The simulated magnetic field can be found from the correspondence (see Ref.~\cite{Lamata2007}):
$\ha_y-\hap_y=\sqrt{2}L(\partial/\partial y)= 2\Delta (\partial/\partial y)$, which gives
$\Delta \Leftrightarrow L/\sqrt{2}$. Since
$c \Leftrightarrow 2\eta\Delta\tilde{\Omega}$ and $mc^2 \Leftrightarrow \hbar \Omega$, we have

\begin{equation}  \label{IonB}
 \kappa = \frac{\hbar eB}{m(2mc^2)} = \left(\frac{\eta\tilde{\Omega}}{\Omega}\right)^2.
\end{equation}

\begin{figure}
\includegraphics[width=8.5cm,height=8.cm]{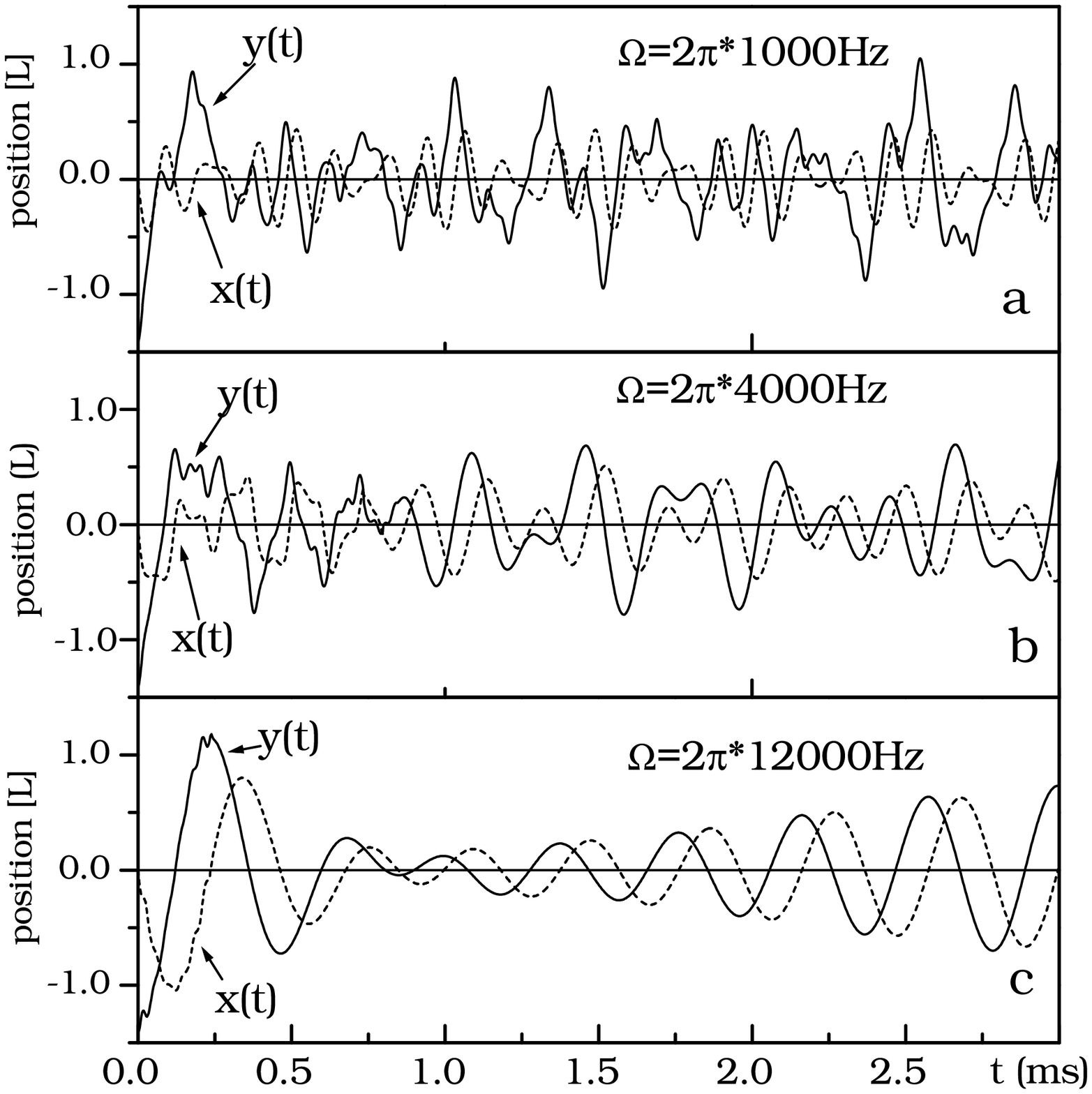}
\caption{Calculated motion of the electron wave packet as described by 2+1 DE,
         simulated by trapped $^{40}$Ca$^{+}$ ions for three values of the effective rest energy $\Omega$.
         Trap parameters: $\eta=0.06$, $\tilde{\Omega}=2\pi\times 68$ kHz, $\Delta\simeq 96\AA$,
         $k_{0x}=\Delta^{-1}$, $d_y=\Delta\sqrt{2}$,
         $d_x=0.9d_y$. Simulations correspond to ratios $\kappa=\hbar\omega_c/2mc^2$=
         16.65(a), 1.05 (b), 0.116 (c), respectively. Positions are given in $L=\sqrt{2}\Delta$.} \label{FigD}
\end{figure}
Therefore, by adjusting frequencies  $\Omega$ and $\tilde{\Omega}$ one can simulate different regimes of
the ratio $\kappa=\hbar\omega_c/2mc^2$.
In Fig.~\ref{FigD} we show the calculated ZB for three values of $\kappa$:
16.65, 1.05, 0.116. It is seen that, as $\kappa$ gets larger (i.e. the field intensity increases or the
effective gap decreases), the frequency spectrum of ZB becomes richer.
This means that more interband and intraband frequency components contribute to the spectrum.
Both intraband and interband frequencies correspond to the selection rules $n'=n\pm 1$ so that,
for example, one deals with ZB (interband) energies between Landau levels $n=0$ to $n'=1$, and
$n=1$ to $n'=0$, as the strongest contributions. For high magnetic fields the interband and intraband
components are comparable, so one can legitimately talk about the ZB.
We believe that ZB oscillations of the type shown in Fig.~\ref{FigD}a, based on the 2+1 DE for
$\kappa=\hbar\omega_c/2mc^2\gg 1$, are the best candidate for an observation of the simulated trembling
motion in the presence of a magnetic field. The calculated spectra shown in Fig.~\ref{FigD} use the
simulated parameters already realized experimentally, see~\cite{Gerritsma2010}.
Notice the tremendous differences of the time and position scales between the results
for free electrons in a vacuum shown in Fig.~\ref{FigC} and the simulated ones in Fig.~\ref{FigD}.

The anisotropy of ZB  with respect to the $\langle x(t)\rangle$
and $\langle y(t)\rangle$ components, seen in Figs. \ref{FigC} and \ref{FigD},
is due to the initial conditions, namely $k_{0x}\neq 0$
and $k_{0y}=0$. A similar anisotropy was predicted in the zero-gap
situation in graphene~\cite{Rusin20008a}.

We emphasize that the Dirac equation (\ref{GenHD}) and our resulting calculation,
as well as Eq.~(\ref{IonHD}) and its simulation in Eq.~(\ref{IonSim}),  represent the 'empty'
Dirac Hamiltonian which does not take into account the 'Fermi sea' of electrons with negative energies
in a vacuum. This one-electron model corresponds to the original considerations of Schrodinger's.
On the other hand, the filled states of electrons with negative energies may affect the
phenomenon of ZB, see~\cite{Krekora2004,Barut1968}.

Finally, one  should recognize that the experiment of Gerritsma {\it et al.}~\cite{Gerritsma2010}
simulates not only the 1+1 Dirac equation for free relativistic electrons in a vacuum but
also the two-band {\bf k.p} model for electrons in narrow-gap semiconductors and the ZB resulting from
this description~\cite{Zawadzki2005KP,Zawadzki2010}. In fact, the results of Ref.~\cite{Gerritsma2010}
look remarkably similar to our predictions~\cite{Zawadzki2010}.

In summary, we calculated the trembling motion of relativistic
electrons in a vacuum in the presence of a magnetic
field for 3+1 and 2+1 spaces. In contrast to the no-field case, the presence of a magnetic
field results in many interband frequencies contributing to the trembling motion. In the 2+1 case
the ZB oscillations of the electron wave packet are stationary, i.e. they do no decay in time. We
indicate how to simulate the Dirac electron in a magnetic field and the resulting ZB using trapped ions
and laser excitations. We show that, for the parameters of DE simulated very recently by
Gerritsma {\it et al.}~\cite{Gerritsma2010},
the effect of a magnetic field on the trembling motion should be clearly observable.

\end{document}